\documentstyle[11pt]{article}

   \newcommand{\be}{\begin{equation}}
   \newcommand{\ee}{\end{equation}}
   \newcommand{\lb}{\label}

   \newcommand{\bv}{{\bf v}}
   \newcommand{\br}{{\bf r}}
   \newcommand{\bk}{{\bf k}}
   \newcommand{\bl}{{\bf l}}

   \newcommand{\bE}{\hat{{\bf e}}}
   
   \newcommand{\bR}{{\bf R}}
   
   \newcommand{\bT}{{\bf T}}
   \newcommand{\bA}{{\bf A}}
   \newcommand{\bU}{{\bf U}}
   \newcommand{\bC}{{\bf C}}
   \newcommand{\bL}{{\bf L}}
   \newcommand{\bI}{{\bf I}}
   
   \newcommand{\bu}{{\bf u}}
   
   \newcommand{\bs}{{\bf s}}

   \newcommand{\bx}{{\bf x}}

   \newcommand{\vl}{\overline{{\bf v}}}
   \newcommand{\vs}{{\bf v}^{\prime}}
   \newcommand{\ul}{\overline{{\bf u}}}
   
   \newcommand{\pl}{\overline{p}}
   
   \newcommand{\btau}{{\mbox{\boldmath $\tau$}}}
   \newcommand{\grad}{{\mbox{\boldmath $\nabla$}}}
   
   \newcommand{\bsigma}{\overline{\mbox{\boldmath $\sigma$}}}

\begin{document}
   \title{ The `Multifractal Model' of Turbulence and {\em A Priori} Estimates in
   Large-Eddy Simulation, II.
	   Evaluation of Stress Models and Non-Universal Effects of the Filter}
   \author{Gregory L. Eyink\\{\em Department of Mathematics, University of
   Arizona}\\{\em Tucson, AZ 85721}}
   \date{ }
   \maketitle
   \begin{abstract}
   We continue a previous work in which {\em a priori} estimates were derived on
   subgrid stress and subgrid flux for filtering schemes used in the turbulence 
   modelling method of Large-Eddy Simulation (LES). The estimates were derived 
   there as rigorous consequences of the exact subgrid stress formulae from 
   Navier-Stokes equations under the conditions assumed for velocity fields in 
   the Parisi-Frisch ``multifractal model.'' It was also shown that these
   assumptions are realistic in an extended inertial range. Therefore the
   estimates must be obeyed by any faithful subgrid model and we use them here 
   to evaluate some popular models of the subgrid stress (Smagorinsky, Bardina, 
   etc.) We also examine the effects of the choice of filter function on the 
   magnitudes of subgrid stress and transfer. Under mild assumptions on the 
   filter  these quantities are determined by local-in-wavenumber, 
   inertial-range interactions and can be modelled in a universal way. However, 
   one common choice of filter---the sharp cutoff filter in Fourier space
   ---does not satisfy the modest required conditions and we show that the 
   associated flux, including ``backscatter'' effects, may be spuriously 
   dominated by nonlocal-in-wavenumber, convective processes of a non-universal 
   type.


   \end{abstract}

   \section{Introduction}

   In the previous part of this work \cite{I1}(hereafter denoted I) we derived
   {\em a priori} estimates on
   subgrid stress and subgrid energy flux as they appear in the filtering approach
   employed in large-eddy
   simulation (LES). The paper I was devoted to more purely theoretical issues.
   The relationship of the
   estimates to Kraichnan's ``refined similarity hypothesis'' for local energy
   flux \cite{Kr74} was also discussed,
   in our earlier work \cite{I-RSH}. It was shown that the estimates arise from
   the fact that LES flux with
   a ``good'' filter, concretely specified by modest conditions, gets its dominant
   contribution in the inertial range
   from local wavevector triads. Thus the traditional picture of the local energy
   cascade was verified and an alternative
   scenario of ``local transfer by nonlocal triads'' suggested by DNS studies
   \cite{DR,OK} was shown to be an artefact
   of the measure of energy flux employed in those works based upon a sharp
   Fourier filter. The sharp cutoff filter
   is the most common example of a ``bad'' filter, which violates the modest
   conditions we imposed. In that case, we
   showed that the associated energy flux may indeed be dominated by distant
   wavevector triads, whose physical origin
   is transfer of energy back and forth across the sharp spectral boundary at
   inertial-range wavenumber $k$ due to rapid
   convection of eddies of size $\approx 1/k$ by energy-range eddies. However, it
   was pointed out that energy flux defined by
   a ``good'' filter is a {\em scale-average} of the sharp spectral flux. This
   leads to cancellation of the nonlocal
   contribution because of detailed energy conservation instantaneously in each
   flow realization, as already established in
   our earlier work for a special example \cite{I-D}. The LES flux with a ``good''
   filter was therefore shown
   to give an appropriate combination of wavevector triads in which convective
   effects to transfer cancel, concretely
   realizing independent proposals of Waleffe \cite{FW} and Zhou \cite{YZ} that
   such a combination should exist. All
   of the previous results were derived based upon the Parisi-Frisch
   ``multifractal model'' of inertial-range velocity fields,
   which assumes local H\"{o}lder exponents $h<1$ distributed throughout the flow
   as the molecular viscosity goes to zero
   \cite{FP,Fr91,Fr95}. It was shown that such a hypothesis is necessary, with
   further $h\leq 1/3$ on some space set, in order
   to maintain constant mean energy flux. Additional rigorous results of
   \cite{I-B} were reviewed which establish that
   the local H\"{o}lder continuity must hold almost surely in every realization
   under very widely accepted assumptions,
   such as power-law decay of the mean energy spectrum.

   In this paper we will discuss the implications of our estimates in I for
   practical LES work. Let us just
   briefly recall the motivation of this approach \cite{WR,UP,MG1}. If a filtering
   function
   $G_\ell(\br)=\ell^{-d}G_\ell(\br/\ell)$ is convoluted with the velocity field,
   as
   \be \vl_\ell(\br)\equiv \int d^d\br'\,\,G_\ell(\br-\br')\bv(\br'), \lb{I1} \ee
   then this provides a convenient definition of the ``large-scale velocity'' or
   ``resolved field'' $\vl_\ell.$
   The ``small-scale velocity'' or ``subgrid field'' can be simply defined as the
   complementary component:
   \be \vs_\ell(\br)\equiv \bv(\br)-\vl_\ell(\br). \lb{I2} \ee
   When the filtering operation is applied to the Navier-Stokes system an equation
   is obtained for $\vl_\ell$:
   \be \partial_t \vl_\ell+\nabla\cdot(\vl_\ell\vl_\ell+\btau_\ell)=-\nabla
   \pl_\ell+\nu_0 \bigtriangleup \vl_\ell, \lb{I3} \ee
   in which $\pl_\ell$ is the filtered pressure field (required to maintain
   $\nabla\cdot\vl_\ell=0$) and $\btau_\ell$
   is the subscale turbulent stress. It is given as an explicit quadratic function
   of the velocity
   \be \btau_\ell=\bT_\ell(\bv,\bv) \lb{I4} \ee
   with
   \be \bT_\ell(\bu,\bv)\equiv \overline{(\bu\bv)}_\ell-\ul_\ell\vl_\ell. \lb{I5}
   \ee
   Because of this last term the Eq.(\ref{I3}) is not a closed equation for
   $\vl_\ell.$ The main effort of LES practitioners is
   to devise suitable model representations $\btau_\ell={\bf T}_\ell^{{\rm
   model}}(\vl_\ell,\vl_\ell)$ as functions of
   $\vl_\ell$ in order to close these equations, allowing them to be solved
   numerically. (It is actually only the trace-free
   or deviatoric  part, $\stackrel{\circ}{\bT}_\ell\equiv
   \bT_\ell-{{1}\over{3}}\left({\rm Tr}\bT_\ell\right)\bI$ which is
   generally modelled, while the diagonal part is absorbed into the pressure
   term.) The theoretical justification for such
   modelling lies in the presumed {\em universality of the small-scales}
   $\vs_\ell$ when the grid-scale parameter $\ell$ is
   chosen as an inertial-range length. In that case, a generally applicable or
   universal model of the stresses may have
   some validity. The computational advantage of LES in very high Reynolds number
   flow is that the filtered
   Eq.(\ref{I3}) may be solved numerically on a computational grid with
   point-separation $\tilde{\Delta}\approx \ell/2$
   rather than with a mesh which resolves scales down to the viscous length
   $\eta,$ as necessary for direct numerical
   simulation (DNS) of Navier-Stokes. It is more customary in the LES literature
   to denote the filtering scale as $\Delta$
   rather than $\ell,$ because of this fact.

   One of our main theses here is that the estimates we have established may be
   used to place some constraints on models
   {\em a priori} since the bounds satisfied for the {\em exact} subscale stress
   must be satisfied also for a
   faithful model. That is, given the true velocity field, with local H\"{o}lder
   regularity in the inertial-range,
   the two functions $\bT_\ell(\bv,\bv)$ and $\bT^{{\rm
   model}}_\ell(\vl_\ell,\vl_\ell)$ should scale in the
   same way at each point $\br$ as $\ell\rightarrow 0$ through that range.
   Although an actual model simulation is performed
   with fixed $\ell=\Delta,$ it seems reasonable to require the previous
   correspondence for $\ell\rightarrow 0$ since it is
   based upon the ideas of universality and scale-similarity which underlie the
   whole scheme. In particular, a model which
   has this property will faithfully reflect ``intermittency'' effects due to the
   range of H\"{o}lder indices
   $h<1$ in the flow associated to structures either more or less singular than
   the K41 mean-field value $h={{1}\over{3}}.$
   This is correct at least for the model stresses and true stresses obtained by
   filtering the {\em true velocities}, if not for
   the model stress obtained from the LES simulation itself compared with the true
   stress. The proposal is similar in nature
   to the so-called {\em a priori test} which was first suggested by Clark et al.
   \cite{CFR} (and sometimes called the ``Clark
   test''). In that case, it was proposed to judge models by their ability to
   agree with the true subscale stress calculated by
   filtering velocity fields obtained as data from a (nominally) high Reynolds
   number DNS or well-resolved experimental
   measurements. It has also the same spirit as the ``statistical {\em a priori}
   tests'' proposed recently by Meneveau
   \cite{CM}. As for both of these proposals it is recognized that satisfaction of
   the {\em a priori} test cannot guarantee
   that a model will perform well for some specific task, such as accurate
   prediction of mean velocities or energies, faithful
   reproduction of large-scale flow structures, etc. On the other hand, if the
   {\em a priori} estimates are invalid for
   a specific model, then it shows that the model will give erroneous predictions
   for the true stress {\em even if the
   filter is applied to true velocities}. Hence errors will be incurred in
   addition to those that arise from the
   discrepancies between the true resolved velocity and that produced by the
   numerical LES. For that reason the satisfaction
   of the {\em a priori} estimates by the model stress is a {\em necessary}
   condition to accurately reproduce the subscale
   stress in a numerical LES, unless some fortuitous cancellation of errors can
   occur between the false scaling of the
   model and the deviations of the true and numerical velocities.

   A second main thesis of this work involves the effects of the filter choice in
   LES work. It was already observed in I
   that very distinctively different results are obtained for magnitudes of energy
   transfer in the case of ``good''
   filters which satisfy our modest criterion and ``bad'' filters which violate
   it. This point has practical importance
   since one commonly used filter, the sharp Fourier cutoff filter, is ``bad''
   according to our criterion. In that case,
   the convective effects discussed in I will be present if the cutoff filter is
   applied to the actual velocities.
   In particular, as we discuss further below, for a ``bad'' filter the distant
   triads produce---in the usual language of LES
   ---a large backscatter of energy which is convective in origin, whereas with a
   ``good''  filter the backscatter is due only
   to local-in-wavenumber straining processes. We therefore draw here an important
   distinction between {\em convective
   backscatter} and {\em strain backscatter}, which does not seem to have been
   emphasized before. Since the convective
   backscatter has its origin in interactions with the large-scale, inhomogeneous
   eddies, it cannot be universally modelled.
   Any attempt to do so will require an elaborate alteration of the model tailored
   to the large-scale motions {\em if it is
   even possible at all.} This type of problem, which is predicted by our rigorous
   {\em a priori} analysis of the equations
   of motion, will be shown to occur in the previous attempts to model
   ``convective sweeping,'' such as that of Chasnov
   \cite{Ch}. Hence we shall argue that modelling of the convection backscatter is
   ill-advised and, practically speaking,
   impossible for inhomogeneous flows. Even for homogeneous turbulence it leads to
   reduced computational economy of the
   simulation without any corresponding increase in its faithfulness or accuracy.
   For a ``good'' choice of filter
   the convective backscatter is entirely absent and hence there is no rationale
   to incorporate it in numerical LES.
   One should keep in mind that in most current LES computations the explicit
   filter function never appears anyway unless a
   ``defiltering'' of the computed velocity is attempted. Therefore, the primary
   practical implication of this work is on
   the choice of filter used for filtering studies of either DNS or numerical data
   to determine what statistics of the subgrid
   stress ought to be used for modelling purposes.

   The precise contents of this work are as follows: in Section 2 we shall derive
   a somewhat more general form of our
   estimates than in I and discuss their validity for specific filters commonly
   employed in LES. Thereafter we shall
   discuss implications of the estimates for LES and, in particular, investigate
   their {\em a priori} validity
   for some common stress models. In Section 3 we shall discuss the issue of
   ``convective vs. strain backscatter''
   based upon our previous work in I and explain why the former must lead to
   non-universal effects. We then
   discuss how such behavior is seen also in analytic closures of EDQNM-type and
   in previous attempts based upon
   them to model ``convective backscatter.'' Morals for LES will be drawn.
   Finally, in the Section 4 we will make our
   conclusions.

   \section{{\em A Priori} Estimates and Evaluation of Stress Models}

   \noindent {\em (2.1) Derivation of the Estimates}

   We shall succinctly rederive and also generalize the estimates in I for
   $\btau_\ell.$ To that end, let us
   review here the {\em Germano decomposition} of the subscale stress \cite{MG2},
   which also forms the most
   rational basis to introduce the common subgrid stress models in the next
   subsection. The idea of the decomposition is
   straightforward. One substitutes $\bv=\vl_\ell+\vs_\ell$ into $\bT_\ell$ to
   obtain:
   \be \bT_\ell(\bv,\bv)=\bL_\ell(\bv,\bv)+\bC_\ell(\bv,\bv)+\bR_\ell(\bv,\bv),
   \lb{II1} \ee
   where
   \be \bL_\ell(\bv,\bv)\equiv \bT_\ell(\vl_\ell,\vl_\ell), \lb{II2} \ee
   is the so-called {\em Leonard stress},
   \be \bC_\ell(\bv,\bv)\equiv
   \bT_\ell(\vl_\ell,\vs_\ell)+\bT_\ell(\vs_\ell,\vl_\ell), \lb{II3} \ee
   is the {\em cross stress}, and
   \be \bR_\ell(\bv,\bv)\equiv\bT_\ell(\vs_\ell,\vs_\ell), \lb{II4} \ee
   is denoted as {\em Reynolds stress}. To be more precise, these are ``modified''
   forms. \footnote{The traditional
   decomposition \cite{AL} was instead
   \be \bT_\ell(\bv,\bv)=\tilde{\bL}_\ell(\bv,\bv)+\tilde{\bC}_\ell(\bv,\bv)
			     +\tilde{\bR}_\ell(\bv,\bv), \lb{II5} \ee
   with
   \be \tilde{\bL}_\ell(\bv,\bv)\equiv\overline{(\vl_\ell\vl_\ell)}_\ell
				      -\vl_\ell\vl_\ell, \lb{II6} \ee
   and
   \be \tilde{\bC}_\ell(\bv,\bv)\equiv\overline{(\vl_\ell\vs_\ell)}_\ell
		    +\overline{(\vs_\ell\vl_\ell)}_\ell, \lb{II7} \ee
   and
   \be \tilde{\bR}_\ell(\bv,\bv)\equiv\overline{(\vs_\ell\vs_\ell)}_\ell. \lb{II8}
   \ee}
   In the Germano decomposition, as its originator himself emphasized \cite{MG2},
   each term is separately Galilei invariant,
   whereas this is not so for the traditional decomposition.

   We now make here a similar---and related---observation: each term {\em
   separately} in the Germano
   decomposition Eq.(\ref{II1}) obeys the {\em a priori} estimate derived in I for
   $\btau_\ell$ alone.
   We recall that the estimates in I were based upon the {\em local H\"{o}lder
   condition} postulated in the
   Parisi-Frisch ``multifractal model'':
   \be |\bv(\br+\bl)-\bv(\br)|=O\left(\ell^h\right), \lb{II9} \ee
   with $0<h<1,$ which is usually denoted $\bv\in C^h(\br).$ The reader should
   refer to I for the theoretical basis of this
   assumption. Under this hypothesis it was proved in I that
   \be \bT_\ell(\bv,\bv)=O\left(\ell^{2h}\right) \lb{II10} \ee
   at the space point $\br.$ Here we shall extend this same estimate to the
   separate terms $\bL_\ell,\bC_\ell,\bR_\ell.$
   This follows as a consequence of two facts.

   First, there is a simple generalization of the identity of Constantin et al.
   \cite{CET} used in I. It reads
   \be \bT_\ell(\bu,\bv)=\langle \Delta\bu(\br)\Delta\bv(\br)\rangle_\ell
			     -\langle
   \Delta\bu(\br)\rangle_\ell\langle\Delta\bv(\br)\rangle_\ell,
   \lb{II11} \ee
   where $\Delta_\bl\bv(\br)\equiv \bv(\br)-\bv(\br-\bl)$ denotes the ``backward
   difference,'' and
   \be \langle f\rangle_\ell\equiv \int d^d\bl\,\,G_\ell(\bl) f(\bl), \lb{II12}
   \ee
   is a ``separation-average'' of a function $f$ of the displacement vector $\bl$
   with respect to the filter function
   $G_\ell.$ The proof of it is the same as in I. We may note also that
   \be \vs_\ell(\br)=\langle \Delta\bv(\br)\rangle \lb{II13} \ee
   and
   \be \grad\vl_\ell(\br)= -\int d^d\bl \,\,(\grad G_\ell)(\bl)\Delta_\bl\bv(\br),
   \lb{II14} \ee
   which were proved in I. The first identity, Eq.(\ref{II11}), has been
   independently discovered by Vreman et al. \cite{VGK},
   who employed it to establish realizability inequalities for the stress tensor.
   Note that this identity also makes the
   Galilei invariance of each separate term obvious, since each is expressed
   entirely in terms of velocity differences.

   The second fact used to extend our estimates is the observation that, if
   $\bv\in C^h(\br),$ then also
   \be \Delta_\bl\vl_\ell(\br)=O\left(\ell^h\right), \lb{II15} \ee
   and
   \be \Delta_\bl\vs_\ell(\br)=O\left(\ell^h\right). \lb{II16} \ee
   A sufficient condition on the filter for these results to hold is
   \be \int d\bx\,\,\left[|G(\bx)|+|\nabla G(\bx)|\right]\cdot|\bx|^{2}<\infty,
   \lb{II17} \ee
   which was the condition for a ``good'' filter set forth in I. It is not
   necessary and may actually be weakened
   to some extent, but it is anyway very mild in practice. Eq.(\ref{II15}) follows
   directly from the identity
   \be \Delta_{\bl}\vl_\ell(\br)=
	      \int
   d\bs\,\,G_\ell(\bs)\left[\Delta_{\bl+\bs}\bv(\br)-\Delta_{\bs}\bv(\br)\right].
   \lb{II18} \ee
   Note that derivation of the estimate Eq.(\ref{II15}) here uses only the weaker
   condition
   $\int d\bx\,\,|G(\bx)|<\infty$ on the filter function. Since
   \be \Delta_\bl\vs_\ell(\br)=\Delta_\bl\bv(\br)-\Delta_\bl\vl_\ell(\br),
   \lb{II19} \ee
   the second estimatee Eq.(\ref{II16}) immediately follows as well.

   Now we can easily see that the same big-$O$ estimates hold for each of
   $\bL_\ell,\bC_\ell,$ and $\bR_\ell$ as
   for the entire tensor $\bT_\ell.$ Because each of the terms in the Germano
   decomposition just comes from
   replacing each $\bv$ in $\bT_\ell(\bv,\bv)$ by a $\vl_\ell$ or a $\vs_\ell,$
   our earlier analysis of I gives the estimate
   $O\left(\ell^{2h}\right)$ separately to each term. Note that the same is not
   true of the terms in the traditional
   decomposition. This is directly related to their lack of Galilei invariance,
   since each term can have a proportionality
   to the {\em amplitude} of the homogeneous velocity component. The validity of
   the estimate implies that each of the
   terms scales as $\sim \vs_\ell\vs_\ell,$ just as suggested by conventional
   heuristics. In particular, it follows
   under our modest filter requirement Eq.(\ref{II17}) that the stress components
   are truly {\em small-scale} quantities,
   locally-determined in scale and without direct dependence upon the large-scale
   modes. It is precisely this fact that
   makes possible a universal model of the stresses, generally applicable to any
   flow. On the other hand, the cross
   and Leonard stresses in the {\em traditional} decomposition scale instead as
   $\vl_\ell\vs_\ell.$ The ``subgrid stress''
   in that case is not truly a subgrid-scale quantity but depends also upon the
   large scales of motion.

   The estimates on the subgrid stress have another interesting implication which
   was pointed out to us by
   U. Piomelli. Since the H\"{o}lder condition must be satisfied in the
   inertial-range of turbulent flow,
   then Eq.(\ref{II10}) gives some justification to the use of 2nd-order accurate
   numerical
   schemes in integrating LES models. In that case the errors are formally
   $O(\ell^2),$ which is asymptotically
   negligible compared to $\btau_\ell\sim \ell^{2h}$ for small $\ell,$ when $h<1.$
   Reynolds had earlier
   made an estimate that $\btau_\ell\sim \ell^{2}$ and argued that a second-order
   scheme would entail errors
   of the same order as the stress itself and, therefore, be inadequate (U.
   Piomelli, private communication).
   However, Reynolds' estimate assumes the validity of a Taylor expansion to
   first-order for the velocity field,
   or $h>1.$ We have seen in I that this, or even a much weaker condition
   $h>{{1}\over{3}},$ is inconsistent
   with constant mean flux of energy in the inertial range. Therefore, it is not a
   valid estimate in the inertial-range
   and second-order schemes should be numerically adequate.

   Let us consider the validity of the condition Eq.(\ref{II17}) for filters
   commonly used in practical LES.
   The {\em Gaussian filter} is a ``good'' example which easily satisfies
   these constraints. It is usually defined in the LES literature as
   \be G(\bx)=\left({{6}\over{\pi}}\right)^{d/2}
		   \exp\left[-{{6|\bx|^2}\over{\pi}}\right]. \lb{II20} \ee
   However, two other common choices do not satisfy our weak condition. The {\em
   tophat filter}
   \be G(\bx)= \left\{ \begin{array}{ll}
		       1 & \mbox{ if $\max_i|x_i|<{{1}\over{2}}$} \cr
		       0 & \mbox{ otherwise}
		       \end{array}
	       \right. \lb{II21} \ee
   is not differentiable. Nevertheless, although our sufficient condition is
   violated, all of
   our estimates in I and here are still valid for the tophat filter. The only
   place where the differentiability
   of the filter was used was in deriving the estimate
   \be \grad\vl_\ell(\br)=O\left(\ell^{h-1}\right). \lb{II21a} \ee
   This still holds for the tophat filter, since
   \be \nabla_i\vl_\ell(\br)=
      {{1}\over{\ell^d}}\left(\prod_{k\neq i}\int_{-\ell/2}^{\ell/2}dh_k\right)
   \left[\bv\left(\br+h_k\bE_k+{{\ell}\over{2}}\bE_i\right)
		    -\bv\left(\br+h_k\bE_k-{{\ell}\over{2}}\bE_i\right)\right]
   \lb{II22} \ee
   with that filter choice. The {\em sharp cutoff filter}
   also violates condition Eq.(\ref{II17}), this time because of a more serious
   problem of slow spatial decay. The
   filter is defined by its Fourier transform
   \be \widehat{G}(\bk)=\left\{ \begin{array}{ll}
		       1 & \mbox{ if $\max_i|k_i|<\pi$} \cr
		       0 & \mbox{ otherwise.}
		       \end{array}
	       \right. \lb{II23} \ee
   In physical space this gives
   \be G(\bx)= 2^d\prod_{i=1}^d {{\sin\pi x_i}\over{x_i}}, \lb{II24} \ee
   which goes as $|\bx|^{-d}$ for large $\bx.$ Because of this exponent-$d$
   power-law decay, the moment condition
   is not satisfied. In fact, we will see by example in the next section that the
   previous bounds for $\btau_\ell$
   may actually be violated for this filter.

   \noindent {\em (2.2) {\em A Priori} Evaluation of Stress Models}

   We are now in a position to introduce and evaluate the common subgrid stress
   models according to the criterion
   discussed in the Introduction.

   The most common models are very easily described on the basis of the Germano
   decomposition.
   Note that the first Leonard stress term is already a function of the resolved
   field $\vl_\ell$ and does not need to
   be modelled. The remaining terms, however, involve $\vs_\ell$ and must be
   modelled. The simplest, and oldest,
   model is the {\em Smagorinsky model} \cite{JS} which takes
   \be {\stackrel{\circ}{\bT}}_\ell\,^{{\rm SM}}(\vl_\ell,\vl_\ell)\equiv
   -2\nu_\ell\bsigma_\ell, \lb{II25} \ee
   where $\nu_\ell(\br)$ is a local ``eddy-viscosity'' determined by the formula
   \be \nu_\ell\equiv \left(C_S\ell\right)^2\sqrt{2\bsigma_\ell^2}, \lb{II26} \ee
   which arises from an energy-balance argument. Within the Germano decomposition,
   this may be thought
   to come from dropping the Leonard term and taking the previous model for the
   $\bC+\bR$ stress terms.
   Exactly the opposite choice is made in the {\em Bardina model} \cite{BFR},
   which drops the cross and
   Reynolds terms and simply keeps the explicit Leonard stress $\bL$:
   \be \bT^{{\rm BD}}_\ell(\vl_\ell,\vl_\ell)\equiv \bT(\vl_\ell,\vl_\ell).
   \lb{II27} \ee
   The most intuitive procedure in these terms is, of course, both to keep $\bL$
   and to model $\bC+\bR$
   \`{a} la Smagorinsky, which yields the {\em mixed model}:
   \be {\stackrel{\circ}{\bT}}_\ell\,^{{\rm MX}}(\vl_\ell,\vl_\ell)\equiv
   {\stackrel{\circ}{\bT}}(\vl_\ell,\vl_\ell)-2\nu_\ell\bsigma_\ell, \lb{II28} \ee
   These may be regarded as the ``classical'' subgrid stress models for LES.

   It is very easy to verify that these models all satisfy the {\em a priori}
   estimates given our prior
   discussion. In fact, since $\bT^{{\rm BD}}_\ell=\bL_\ell$ identically, the
   estimates for the Bardina model follow
   from our discussion above for the Leonard stress. In the case of the
   Smagorinsky model we infer from Eq.(\ref{II21a}) that
   \be \bsigma_\ell=O\left(\ell^{h-1}\right), \lb{II29} \ee
   which gives also
   \be \nu_\ell=O\left(\ell^{h+1}\right) \lb{II30} \ee
   from the prescription Eq.(\ref{II26}) for the local eddy viscosity. Therefore,
   it follows that
   \be {\stackrel{\circ}{\bT}}_\ell\,^{{\rm SM}}=O\left(\ell^{2h}\right),
   \lb{II31} \ee
   as required. Finally, the mixed model is just the sum of the previous two, so
   it obeys the estimates as well.
   As a consequence, all of these models will, if evaluated for the true velocity
   fields in the inertial range,
   obey the same big-O estimates as the true subgrid stress at the same space
   point. This gives one some hope
   that the models will be able to capture ``intermittency'' effects due to
   structures in the flow.

   Let us discuss some other models besides the above ``standard'' models.
   A rather recent proposal, made and implemented by Germano et al. \cite{MG3},
   attempts to improve upon the
   Smagorinsky and mixed models by using information from the evolving flow itself
   to fix the parameter
   $C_S,$ which is no longer a constant but ranges over both positive and negative
   real values. This is the
   so-called {\em dynamical model}. We shall not discuss all the versions of this
   model---of which there
   are several in the literature---or its precise derivation. In none of the
   published cases have we been able to prove
   the {\em a priori} estimates. However, this is not necessarily a negative
   conclusion. Merely having unwieldy
   expressions for which it is inconvenient to establish a mathematical estimate
   is no reason to disqualify a model.
   Rather it must be shown that the estimates are definitely violated. For
   example, it may be possible to verify
   the estimates {\em a posteriori}. In the case of the dynamical model the
   estimates {\em will} be satisfied
   according to our previous analysis if the model parameter $C_S,$ while not
   constant, stays between fixed
   bounds as $\ell$ is decreased. In the work of Zang et al. on a lid-driven
   cavity flow \cite{YZl}, the parameter
   $C_S$ was observed in the midplane of the flow to range from -0.12 to 0.1 for a
   Smagorinsky-based
   model and from -0.018 to 0.026 for a dynamical mixed model. If these ranges of
   values were observed not to
   increase under successive refinement of the grid, then the estimates would be
   {\em a posteriori} verified.
   This, then, is an important question to consider numerically.

   So far we have not discussed any ``bad'' models. Some of the earlier models
   based upon the traditional
   Leonard decomposition of the subscale stress, e.g. the model used by Moin  and
   Kim in an early LES study
   of turbulent channel flow \cite{MK}, have problems in this regard. We shall not
   discuss this in detail, since
   these models are no longer in common use, but we shall just mention that the
   problems with the estimates in these cases
   are directly connected with the lack of Galilei invariance discussed by
   Germano. However, there are more subtle things
   that can go wrong. All of the estimates verified above depended upon using a
   ``good'' filter that satisfies
   a modest condition like Eq.(\ref{II17}). Nevertheless, there are ``bad''
   filters for which the {\em a priori} estimates
   are not satisfied, including one popular choice, the sharp spectral cutoff. We
   now discuss some of the problems this entails.

   \section{The Fourier Cutoff Filter and Non-Universal Effects}

   \noindent {\em (3.1) Violation of the Stress Estimates}

   We have already observed in I that the energy flux defined with the sharp
   spectral filter
   does not satisfy our {\em a priori} estimates because it is dominated
   instantaneously by
   nonlocal, convective interactions. It is not enough to observe that the
   condition Eq.(\ref{II17}) is
   violated, because it is only a sufficient condition and not a necessary one.
   Instead, the estimates must be directly
   verified to fail. For that purpose we used an example from \cite{I-D}, of the
   form
   \begin{eqnarray}
   \bv(\br)&\equiv& 2U(\bE_1-\bE_2)\sin\left[(\bE_1+\bE_2)\cdot\br\right] \cr
	  \,& &+\sum_{N\geq 1}
   {{2A}\over{2^{Nh}}}\bE_3\left\{\cos\left[2^N\bE_1\cdot\br\right]
   +\cos\left[(2^{N+1}-1)\bE_1\cdot\br-\bE_2\cdot\br\right]\right\}, \lb{III1}
   \end{eqnarray}
   with $U,A$ real constants and $0<h<1.$ This velocity field is given by a
   so-called Fourier-Weierstrass series
   and it is known that $\bv\in C^h(\br)$ for every space point $\br$ in the
   periodic box where it is defined.
   We will just cite the result obtained in \cite{I-D}, without repeating the
   calculation, that
   \be \Pi_\Lambda=\left(2^{2-h}UA^2\right)\times 2^{\Lambda(1-2h)} \lb{III2} \ee
   where $\Pi_\Lambda$ is the (global) subgrid energy flux across wavenumber
   $2^\Lambda$ defined with the sharp
   cutoff filter. Observe the very important feature that the flux is here
   proportional to the
   amplitude $U$ of the lowest excited wavenumber mode. Note that this is not due
   to any failure
   of Galilei invariance. The spectral flux $\Pi(k)$ for $k\neq 0$ is perfectly
   invariant to changes in the
   zero-wavenumber amplitude. However, its leading term  may still have a direct
   dependence upon the
   amplitude of very small wavenumber modes for arbitrarily large $k.$ The physics
   of this, as
   we discussed previously, is the back and forth transfer of energy across the
   spectral boundary $k$ by
   small steps, associated to the rapid convection of small structures of scale
   $\sim 1/k$ by the
   largest eddies. In the language of LES modelling there will be a large {\em
   backscatter}
   due to this effect, which we may call {\em convective backscatter.}

   It might appear that the example we have considered is rather contrived.
   However, a little thought
   shows that, so long as one keeps a sharp spectral cutoff, the same types of
   effects will occur {\em generically}
   in $\Pi_\ell.$ In fact, if a given velocity field had instantaneous transfer
   determined by local triads, its
   superposition with the above velocity field will have transfer dominated by the
   convective, distant triad
   interactions, which are asymptotically much larger. The physical interpretation
   also argues that such effects
   will be quite typical. However, we have seen that this type of process---while
   a real physical effect---is
   essentially spurious in terms of energy transfer, since it averages away over
   time and cancels instantaneously in
   sensitive measures of the flux. In particular, we showed in I that this
   contribution from distant triads cancels
   in LES flux with a ``good'' filter, due to detailed energy conservation. The
   reader is encouraged to review
   Section 3.1 of I where this cancellation of the convective contribution was
   demonstrated.
   It must be distinguished from the backscatter which arises from the local,
   straining
   processes whose contribution to energy flux in 3D is forward on
   average---because stretching of
   isotropic distributions of small-scale vortex elements by large-scale strain
   increases their
   mean energy---but which also has both negative and positive fluctuating values.
   Since these local, stretching
   interactions are the crucial ones for energy transfer, it is this second
   backscatter, or
   {\em strain backscatter} which is more meaningful to the dynamics.
   Nevertheless, the ``convective backscatter''
   has, as we have seen, a much bigger absolute magnitude than the ``strain
   backscatter'' since
   it is nonlocally driven by the largest, most energetic eddies. Therefore, if an
   inappropriate filter
   is used---such as the sharp Fourier cutoff---then the ``convective
   backscatter'' will completely
   dominate and mask the more dynamically intrinsic ``strain backscatter.''

   This seems to be exactly the effect observed in numerical simulations of
   Piomelli et al. \cite{UPl}.  It was
   found in that work that with the sharp cutoff filter the fraction of space
   points instantaneously experiencing
   backscatter was about 50\%,  compared with about 30\% for the Gaussian filter
   in the same flow.
   Furthermore, the intensity of backscatter with the sharp Fourier filter was as
   much as an
   order of magnitude higher than for the Gaussian. Both of these facts are
   consistent with the
   analyses we have made and point toward large convective effects in the sharp
   cutoff case. It was
   independently argued by Vreman et al. \cite{VGK} that the non-positivity of the
   cutoff filter might increase
   the fraction of space points with negative flux. This argument is not so
   clearly valid, since the flux
   is not positive even with a positive filter and can take either sign. Our
   explanation is rather that
   the large convective contributions to the flux of {\em both} signs, positive
   and negative, are present with
   the cutoff filter. Since the convection effects will cancel on average even for
   the cutoff filter, one
   should expect that the fraction of points with either sign of flux will be
   roughly $50\%-50\%$ and that the overall
   magnitude will be much higher. Both of these effects are observed. Notice that
   our explanation of this effect has predictive power. In fact, if the large
   backscatter observed in \cite{UPl}
   is ``convective'' in origin, as we have proposed, then it ought to be absent if
   the same study
   were performed for Kraichnan's ``modified Navier-Stokes system'' (Section 6 of
   \cite{Kr64}.)
   That dynamics lacks the nonlocal, convective interactions which we have
   proposed as responsible for
   the effect, so that the results obtained there for the Gaussian and the sharp
   Fourier filters
   ought to be essentially indistinguishable. This is a crucial test that ought to
   be performed.

   Even if our interpretation is accepted as correct, it might be thought by some
   that there is
   nothing intrinsically wrong with the sharp Fourier cutoff. Some account must be
   taken of the
   difference in physics for the ``good'' and ``bad'' filter cases---exemplified
   by Gaussian and
   sharp Fourier, respectively---but a subgrid model with either of these filters
   could
   be constructed so long as it were consistent with the processes that actually
   occur with that filter.
   Since the ``convective backscatter'' actually occurs with the sharp cutoff
   filter, a faithful
   subgrid model should be constructed to represent it. However, we emphatically
   {\em do not agree}
   with such an assessment. It is crucial to understand the source of the
   convection dependence. If one uses the sharp
   Fourier filter, then the local flux is still represented by the formula
   $\Pi_\ell=-\nabla\vl_\ell:\btau_\ell$
   just as for any filter. However, it is not too hard to show that
   $\|\nabla\vl_\ell\|_\infty=O\left(\ell^{h-1}\right)$
   here just as for the ``good'' filter case. (For example, it follows from the
   arguments that
   establish the ``Littlewood-Paley-type'' conditions for H\"{o}lder continuity:
   see \cite{I-D}.) Therefore,
   {\em  the violation of the estimate for the energy flux with the sharp spectral
   cutoff is due to a violation of
   the estimate for the ``subscale stress.''} In other words, the nonlocal,
   convective type interactions must appear
   in this ``subscale stress'' so that it is $\sim \ell^h$ rather than
   $\sim\ell^{2h}.$ The expected estimate
   $\btau_\ell\sim \vs_\ell\vs_\ell$ is incorrect and instead $\btau_\ell\sim
   \vl_\ell\vs_\ell.$

   Such anomalous behavior is explicitly verified for the example above, as we now
   show.
   Our calculation will be done with ${{2\pi}\over{\ell}}=k$ chosen to lie in the
   interval
   $\left( \sqrt{(2^{\Lambda+1}-1)^2+1},2^{\Lambda+1}\right)$ between the top mode
   of $S_\Lambda$
   and the bottom mode of $S_{\Lambda+1}.$ To simplify formulas, we introduce some
   notation, as
   follows:
   \be \bU\equiv U(\bE_1-\bE_2),\,\,\,\,\bA\equiv A\bE_3, \lb{III3} \ee
   \be \varphi\equiv (\bE_1+\bE_2)\cdot\br,\,\,\,\,\theta_{N,1}\equiv
   2^N\bE_1\cdot\br,\,\,\,\,\theta_{N,2}\equiv
   (2^{N+1}-1)\bE_1\cdot\br-\bE_2\cdot\br.
      \lb{III4} \ee
   Note that
   \be \vl_\ell(\br)=2\bU\sin\varphi+2\bA\sum_{N=1}^\Lambda
   2^{-Nh}[\cos\theta_{N,1}+\cos\theta_{N,2}]. \lb{III4a} \ee
   It then follows from an elementary but slightly tedious calculation that
   \begin{eqnarray}
   \, & & \bL_\ell=-(\vl_\ell\vl_\ell)_\ell'= \cr
   \, & &\,\,\,\,\,\,\,\,\,\,\,\,\,\,\,\,\,\,\,\,
	 -2(\bU\bA+\bA\bU)\cdot 2^{-\Lambda h}\sin\theta_{\Lambda+1,1}\cr
   \, & &\,\,\,\,\,\,\,\,\,\,\,\,\,\,\,\,\,\,\,\,
	 -2\bA\bA\sum_{N=1}^\Lambda 2^{-(N+\Lambda)h}
   \left[2\cos(\theta_{\Lambda,2}+\theta_{N,1})+\cos(\theta_{\Lambda,2}
				+\theta_{N,2})\right] \cr
   \, & &\,\,\,\,\,\,\,\,\,\,\,\,\,\,\,\,\,\,\,\,
	 -4\bA\bA\cdot 2^{-2\Lambda h}\cos\theta_{\Lambda+1,1}, \lb{III5}
   \end{eqnarray}
   and
   \begin{eqnarray}
   \, & &
   \bC_\ell=\overline{(\vl_\ell\vs_\ell)}_\ell+\overline{(\vs_\ell\vl_\ell)}_\ell=
   \cr
   \, & &\,\,\,\,\,\,\,\,\,\,\,\,\,\,\,\,\,\,\,\,
	 -2(\bU\bA+\bA\bU)\cdot 2^{-(\Lambda+1)h}\sin\theta_{\Lambda,2} \cr
   \, & &\,\,\,\,\,\,\,\,\,\,\,\,\,\,\,\,\,\,\,\,
	 +2\bA\bA\sum_{N=1}^\Lambda 2^{-(N+\Lambda+1)h}
   \left[2\cos(\theta_{\Lambda+1,1}-\theta_{N,2})+\cos(\theta_{\Lambda+1,1}
			    -\theta_{N,1})\right], \lb{III6}
   \end{eqnarray}
   and
   \begin{eqnarray}
   \, & & \bR_\ell=\overline{(\vs_\ell\vs_\ell)}_\ell= \cr
   \, & &\,\,\,\,\,\,\,\,\,\,\,\,\,\,\,\,\,\,\,\,
	 +4\bA\bA\sum_{N=\Lambda+2}^{\infty} 2^{-(2N-1)h}\cdot\cos\varphi \cr
   \, & &\,\,\,\,\,\,\,\,\,\,\,\,\,\,\,\,\,\,\,\,
	 +4\bA\bA\sum_{N=\Lambda+1}^{\infty} 2^{-2Nh}. \lb{III7}
   \end{eqnarray}
   As one would expect, the Reynolds stress $\bR_\ell$ is here $O(2^{-2\Lambda
   h}),$ the naive result. However, the
   Leonard and cross stress terms, $\bL_\ell$ and $\bC_\ell,$ are not, but give
   contributions $\sim \vl_\ell\vs_\ell.$
   In fact, the dominant terms are ones $\sim U\cdot A2^{-\Lambda h},$
   proportional to the amplitude of the lowest excited
   mode. What the estimates mean is that the ``subscale stress'' for the cutoff
   filter
   is not a true function of the small-scales alone---as one would naively
   believe---but there is a
   residual, implicit dependence upon the large-scales. In principle there is no
   difficulty with the Leonard term,
   because it does not need to be modelled and can be calculated  explicitly.
   However, the cross term is a real problem.

   This fact has severe negative implications for LES modelling with such a
   ``bad'' filter as the
   sharp Fourier one. The entire justification for the LES scheme is the
   universality of small-scales
   in the inertial-range, which allows a model description of the subscale stress
   to be attempted. However, if
   the ``subscale stress'' for a particular choice of filter retains a dependence
   upon the low-wavenumber,
   or large-eddy, degrees of freedom that are highly nonuniversal, anisotropic,
   etc. then we see absolutely
   no reason to believe that a simple, general model could be constructed.  One
   may ask how this should
   affect actual LES practice. After all, in most current LES computations the
   explicit filter function
   never appears anyway unless a ``defiltering'' of the computed velocity is
   attempted. \footnote{However,
   this is {\em not} the case in some more sophisticated approaches, such as the
   ``dynamical model,'' where the
   filtering function appears explicitly. Also, on theoretical grounds, filtering
   the velocity at successive
   time-steps in the course of the computation {\em ought} to be done in LES
   computation, with a filter length
   $\ell$ at least as large as $2\times\Delta,$ the grid-length of the numerical
   discretization. This would guarantee
   that the computed velocity field represents at every step a properly filtered
   object and does not develop
   unrealistically large energy in the grid-scale modes. Needless to say, this
   prescription is not usually adopted
   because available numerical power makes it impractical.} Therefore, the primary
   practical implication of this
   work is on the choice of filter used for filtering studies of either DNS or
   numerical data to determine the
   statistics of the subgrid stress for modelling purposes. {\em Indeed, the key
   implication of our analysis
   is that the large backscatter observed with the Fourier filter in studies such
   as \cite{UPl} ought not to be incorporated
   into LES models, since it is just an artefact of that pathological filter
   choice.} It is not even
   {\em possible} to model such backscatter on the basis of Kolmogorov
   inertial-range ideas, since its origin is
   nonlocal and is dominated by non-universal, low-wavenumber modes. To clarify
   these points, we consider now
   some specific models.

   \noindent{\em (3.2) Evaluation of Models Based on the Spectral Cutoff Filter}

   All of the problems discussed appear in the stochastic LES model developed by
   Chasnov \cite{Ch}, who used the
   sharp spectral filter in conjunction with a Langevin model of the EDQNM closure
   equations.
   {}From our foregoing discussion it follows that---in so far as the Langevin
   model is consistent with reality
   (the Navier-Stokes equations)---use of the sharp Fourier filter will lead to
   forward (eddy viscosity) and backward
   (backscatter) contributions to energy transfer which {\em separately} contain
   the strong effects of convection by
   energy-scale eddies that only cancel in the net flux. In fact, Chasnov observed
   exactly this in his model and, indeed,
   claimed it as a {\em virtue} of his approach that he modelled ``the random
   sweeping of small scales by large scales.''
   We do not agree with his appraisal. Our work has shown that this type of effect
   is not present if a ``good'' filter is
   employed for Navier-Stokes dynamics and need not then be modelled. We should
   stress that we do {\em not} disagree with
   Chasnov on the issue of stochastic modelling of backscatter. Even with a
   ``good'' filter what we have termed ``strain
   backscatter'' will still be present and represent an important part of the
   transfer process. It is only
   the ``convective backscatter'' which we have proved unnecessary to model. It is
   clearly
   preferable from the point of computing economy to dispense with modelling such
   effects which are just artefacts
   of a ``bad'' filter choice. Most importantly, any attempt to model such effects
   severely reduces the universality
   of a model, since it requires explicit consideration of the energy-containing
   eddies, whose statistics
   are not universal. For example, the transfer functions $\eta^+(k|k_m)$ and
   $F^+(k|k_m)$ obtained and plotted
   by Chasnov \cite{Ch} (see his Fig.1 for EDQNM results and Figs.3,5 for LES)
   {\em cannot be transferred from
   one simulation to another without regard of the energy-scale statistics}. Even
   such a trivial change as
   adding a homogeneous (zero-wavenumber) component to the velocity field with
   independent Gaussian statistics
   of large variance $v_0$ will drastically increase the size of the ``spikes'' in
   these functions at the
   cutoff wavenumber $k_m.$ While it possible to laboriously incorporate such
   effects with use of the sharp
   Fourier filter, it is totally unnecessary to do so when using using a sensible
   ``good'' filter.

   On the other hand, no negative conclusions can be drawn from our work regarding
   the ``spectral LES'' methods,
   such as those of Chollet and Lesieur \cite{CL}, M\'{e}tais and Lesieur
   \cite{ML}, and others. In these
   LES schemes, a Fourier-Galerkin truncation is used rather than a
   grid-discretization, the stress model is
   a simple $k$-dependent eddy-viscosity model, and the model dynamics is solved
   by a spectral code. The functional
   form of the eddy-viscosity in $k$ is generally taken from the results of
   analytical closures, such as Kraichnan's
   fundamental work on the TFM equations \cite{Kr76}. Along with Chasnov's, these
   LES models have a somewhat different
   status than the ones we have analyzed above, based upon smooth filtering in
   space and a grid-discretization. In the latter
   schemes we found that the stresses in the common {\em models} (Smagorinsky,
   Bardina, etc.) will correctly reflect the
   magnitude of the {\em true} stresses obtained when a ``good'' filter is applied
   to the individual flow fields (see Eqs.
   (\ref{II17}),(\ref{II31}).) What we can definitely state regarding the
   ``spectral LES''  technique is that its stress model
   via the $k$-dependent eddy-viscosity cannot be regarded as an accurate
   representation---realization by realization
   ---of the {\em true} stress obtained by applying the sharp spectral filter to
   individual velocity fields in the
   turbulent ensemble. In fact, we have seen that the latter operation gives rise
   to a subgrid stress which is dominated
   by distant triads in individual realizations, whose effects only cancel when
   additional (scale- or ensemble-) averaging
   is performed. On the contrary, the eddy-viscosity models from the closures
   incorporate only the effects of the local
   interactions on transfer across the cutoff wavenumber and do not include the
   nonlocal transfers discussed above, which
   are actually present in individual realizations with the sharp Fourier filter!
   This does not rule out that such spectral-LES
   schemes will give good results, particularly for statistical averages, as has
   been often found in practice.

   On the other hand, by examination of the expressions for damping and
   backscatter in Chasnov's work \cite{Ch}, $\eta(k|k_m)$
   and $F(k|k_m),$ given by his Eqs. (22) and (23), we see that the diverging,
   convective contributions {\em are} contained
   in these separate terms. This arises precisely from the range of small $q$ in
   the wavenumber integrations there, i.e.
   $q\leq k_0$ and $p\leq k+k_0.$ One expects from our analysis of the true
   Navier-Stokes dynamics that such contributions
   indeed ought to appear with the sharp Fourier filter but that their net
   contribution to {\em mean} energy transfer will
   cancel. This was explicitly verified by Chasnov through an asymptotic analysis
   of his equations (see Eqs.(20),(21) in
   \cite{Ch}.) Hence, his stochastic model provides a really more faithful account
   of the behavior in individual
   realizations with the sharp Fourier filter than the pure viscosity-type models,
   as \cite{CL,ML}. However, we
   argue that this makes the model {\em less useful} not more so, because it is
   less economical to compute and very
   fundamentally restricted in its universality. Even for the idealized situation
   of homogeneous turbulence
   the basic quantities in the model must be recomputed for every new choice of
   low-wavenumber driving mechanism
   which leads to altered statistics of the energy-range eddies. All of these
   problems are totally eliminated
   by the simple expedient of using a ``good'' filter as defined by our criterion
   Eq.(\ref{II17}).

   We should emphasize again that the verb {\em ``use''} has here a subtle
   interpretation, since in many LES
   simulations the filter is only implicit. It is often argued that the cutoff
   filter should be used to
   generate resolved fields that mimic those from LES with a spectral calculation,
   while the top-hat
   filter is more appropriate for a finite-difference scheme. However, this
   correspondence between {\em numerical
   method} and corresponding {\em filter type} is not at all exact \cite{VS}.
   Hence, whatever the numerical scheme,
   one is better off to employ a stress model without ``convective backscatter.''
   This can always be justified
   on the grounds that the implicit filter is a ``good'' one for which those
   effects are absent according
   to our demonstrations above. The prohibition against the ``bad'' filters, like
   the Fourier cutoff, is more
   direct if the filter is explicit in the model, as for the Bardina model
   generally or for the dynamic model.
   In that case, use of the sharp Fourier filter will require for consistency that
   the convective effects
   be included, with all the problems that entails. Coupled with the difficulty of
   using spectral codes at all in more
   complex, inhomogeneous geometries---which is a current trend of LES research---
   we see the role of the cutoff filter
   in the future as rather restricted.

   \section{Conclusions}

   Already from our work in I we can see that the numerical LES modelling
   technique has a great deal of {\em a priori}
   theoretical support. LES is mathematically justified even for the ``weak
   solutions'' of Euler equations necessary to
   describe non-differentiable ``multifractal'' velocity-fields in the inertial
   range of scales. Furthermore, many of the most
   common stress models satisfy {\em a priori} estimates obeyed by the true stress
   and are able to capture some of the basic
   features of inertial-range intermittency. When a ``good'' choice of filter
   function is made, the subscale stress in the
   filtering approach is a function solely of the small scales and may be
   plausibly modelled in a universal way. However,
   if a ``bad'' filter is used---of which the cutoff filter is the most prominent
   example---then the ``subgrid stress''
   is infected with dependence upon the large-scale modes and can no longer be
   represented in a universal form.

   One caveat concerning the LES method is that it is only really justifiable to
   use when an extended inertial interval
   exists. In \cite{BW} Brasseur and Wei have made an estimate of the Taylor-scale
   Reynolds number
   required to produce $\Delta$ decades of inertial range, as
   \be {\rm Re}_\lambda\sim 400\times 10^{2\Delta/3}. \lb{132} \ee
   In particular, for even a single decade, ${\rm Re}_\lambda\sim 400$ must be
   achieved.
   If this pessimistic assessment is correct, then the LES method may only begin
   to have a sound
   theoretical basis in circumstances of exceedingly high Reynolds number not
   often encountered.
   Of course, there is nothing to prevent the application of the LES method in
   situations where
   its theoretical justification is weak and, indeed, it may do rather well there.
   It would not be the
   first time that the performance of a practical numerical method exceeded the
   parameters of its derivation.

   {\bf Acknowledgements.} I am very grateful to U. Frisch, R. H. Kraichnan, U.
   Piomelli, and Z.-S. She
   for conversations on various problems discussed in this work, as well as the
   LES group at
   CTR and two anonymous referees for some constructive remarks. I also wish to
   thank Weinan E for
   pointing out to me the key identity Eq.(\ref{II11}). This research was funded
   through the NSF
   Grant No. DMR-93-14938, of Y. Oono and N. Goldenfeld, and the support of the
   Physics Department
   of the University of Illinois Urbana-Champaign, where this work was
   accomplished.

   \end{document}